\documentclass{vgtc}                          

\graphicspath{{figures/}{pictures/}{images/}{./}} 

\usepackage{times}                     

\usepackage{tabu}                      
\usepackage{booktabs}                  
\usepackage{lipsum}                    
\usepackage{mwe}                       
\usepackage{soul}
\usepackage{xcolor}
\usepackage{mathptmx}                  

\onlineid{0}

\vgtccategory{Research}

\vgtcinsertpkg

\title{Digital Bricolage: Design Speculations for Embodied Approaches to Digitized Print-based Cultural Collections}

\author{
    Malak Sadek\thanks{mfzas2@cam.ac.uk} \\ \parbox{1.4in}{\scriptsize \centering Centre for Human-Inspired Artificial Intelligence (CHIA), Cambridge University, Cambridge, UK.}
    \and Loraine Clarke\thanks{lec24@st-andrews.ac.uk} \\ \parbox{1.4in}{\scriptsize \centering School of Computer Science, University of St Andrews, St Andrews, UK.}
    \and Stefania Forlini\thanks{sforlini@ucalgary.ca} \\ \parbox{1.4in}{\scriptsize \centering Department of English, University of Calgary, Calgary, Alberta, Canada.}
    \and Uta Hinrichs\thanks{uhinrich@ed.ac.uk} \\  \parbox{1.4in}{\scriptsize \centering School of Informatics, University of Edinburgh, Edinburgh, UK.}}
        
\teaser{
  \centering
  \includegraphics[width=\linewidth]{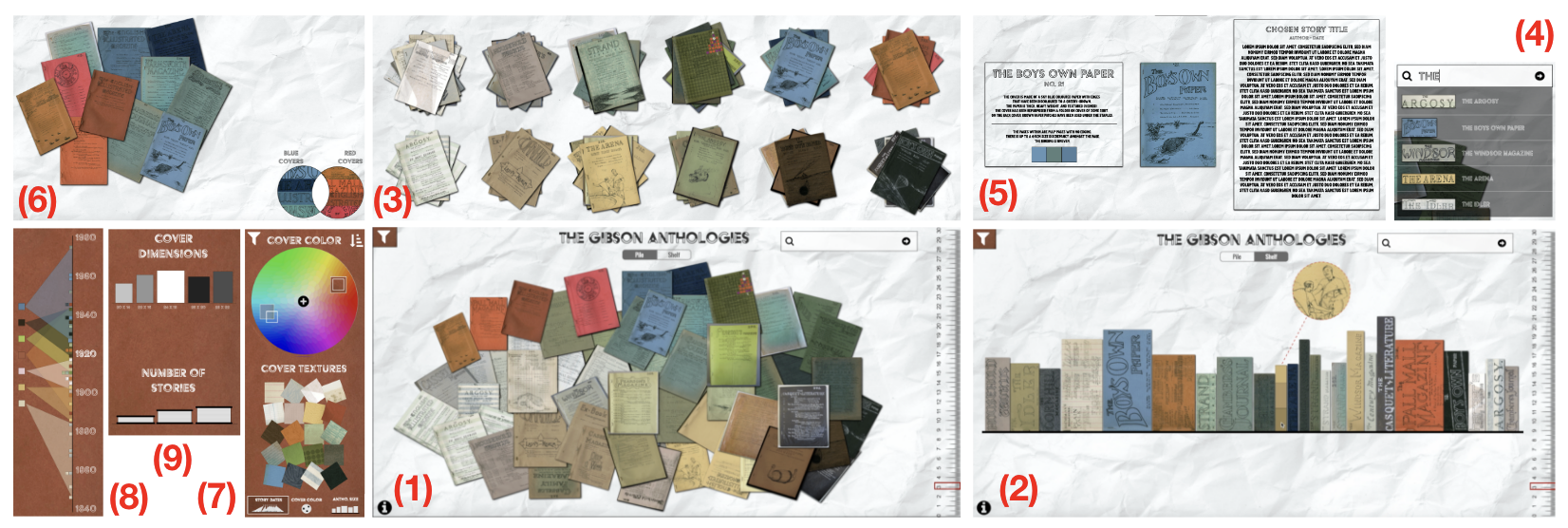}
  \caption{Digital Bricolage features a collection overview that allows toggling between the ``pile'' (1) or ``book shelf'' layout (2). Items can also be clustered based on colour (3). The search bar highlights items' unique typography (4). The detail view on items combines unique bibliographic and material information (5). Items can be filtered by colour(s) (6), cover texture (7), time (8), or physical dimensions (width, height, thickness or page count (9).}
  \label{fig:teaser}
}

\abstract{
  COVID-related closures of public and academic libraries have underlined the importance of online platforms that provide access to digitized print-based collections. However, they also have highlighted the value of in-person handling of print artefacts for sensing and making sense of them. How do existing dominant digital platforms invite and/or discourage embodied forms of exploration and sense-making? What opportunities for embodied experience might we discover if we embrace the material qualities of print-based collections when designing interfaces for digital access? In this paper, we present findings from a speculative exercise where we invited creative professionals and experts in curating and handling access to collections to reflect on existing approaches to digitized print-based collections and to speculate about alternative design opportunities and modes of engagement. We argue for digital bricolage—a design approach that values working with materials that are “on hand” and embracing our ability to “handle” them in ways that foster both casual and curious exploration.
} 


\begin{document}


\firstsection{Introduction}

\maketitle

Material culture studies, history of the book, media theorists and literary critics have long pointed out that content cannot be separated from format, and that a print artifact and its digital copy are fundamentally distinct objects~\cite{drucker2013,digitalart}, not to be treated interchangeably~\cite{drucker2002}. Rather, as highlighted by Manoff, we should recognize the ways in which \textit{``knowledge is shaped by the means through which it is developed, organized, and transmitted''}~[p.322]~\cite{manoff2006}. Similarly, while \textit{``interfaces promote illusions of transparency,''} they themselves \textit{``are not transparent windows''}~[p.320]~\cite{manoff2006}. Instead, they fundamentally shape our experience and understanding of the online environment, the digital copies displayed therein, and any research conducted with such resources. 

Mass digitization has improved the accessibility of collections globally and allows faceted exploration where individual items can co-exist within different search results, making it easier to understand relationships, look for related materials, and get a sense of the broader context of a collection~\cite{facet, visualizing}. It has also made the exploration and analysis of large-scale collections possible, leading to cultural and historical discoveries (e.g.,~\cite{mandala, specpractices}) that would not have been possible manually. However, the focus on semantic content, with little to no concern for the look-and-feel of print artifacts' physical form and format has exasperated the limitations of current digitization, electronic display practices, and resulting digital libraries in which digital ``copies'' or ``surrogates'' of print artifacts are wrongly considered interchangeable and static artifacts. As a result, much of the print artifacts' material characteristics are often obscured or lost. At the same time, interfaces to digital libraries heavily constrain the engagement with collections to canonical search and filtering mechanisms. Nowviskie has highlighted this as a design problem: \textit{``[w]e’re building our digital libraries to be received by audiences as lenses for retrospect, rather than as stages to be leapt upon by performers, co-creators''}~\cite{nowviskie2016}. Insisting that cultural heritage collections \textit{``will never be something we passively encounter,''} Nowviskie challenges us to design interfaces for what she calls \textit{``speculative collections''}---collections that activate imaginations, that function as \textit{``improv spaces''} and frame cultural heritage collections as \textit{``playable''} instruments, things that can be actively used as tools by communities to repair past harm and fashion alternative futures~\cite{nowviskie2016}. 

Nowviskie re-orients concerns about preservation and display of cultural heritage toward inviting active engagement with such materials as a way to catalyze repair and creation. This raises a number of questions: \textit{\textbf{What would it mean to imagine digital collections as performance spaces? How do dominant digital platforms invite and/or discourage embodied forms of exploration and sense-making?}} And: \textit{\textbf{What opportunities for embodied experience might we discover if we embrace the material qualities of print-based collections when designing for digital access?}} 

We present \textit{Digital Bricolage} as a visualization design study conducted during the pandemic, through which we sought to explore ways to increase appreciation of physical artifacts in digital environments. Digital Bricolage can be considered a speculative approach to re-thinking modes of representing cultural print collections, allowing us to experiment with ways to highlight and celebrate the materiality of print collections. We do not believe it is possible to digitally replicate the experience of physically interacting with print artifacts. Yet, we sought to accentuate what digital displays often obscure: the unique, physical formats of print artifacts as ``scenes of evidence'', rich with information about the specific time, place, and modes of their production~\cite{stauffer2012}. We presented Digital Bricolage as an experimental prototype to 23~creative professionals and print collection experts, to engage in a speculative dialogue around the ideas embodied in Digital Bricolage, in comparison to canonical interfaces to digital collections. Digital Bricolage provoked discussions around the potential of object-oriented representations and spatial re-orientation to (1)~support serendipity through generosity and multiple entry points, (2)~emphasize conceptual relationships between items, (3)~offer users more agency over their curatorial experience, and (4)~foreground tactile handling and a link to both physical spaces and dimensionality.

Through Digital Bricolage we hope to provoke constructive discussions about GLAM interfaces that can move away from pragmatic grid-based, ``one-fits-all'' solutions toward design approaches that emphasize collection objects as unique artifacts and foreground embodied engagement, while evoking a sense of agency, creativity, and productive speculation. We call for moving beyond visual generosity, foregrounding collections' interface biases, assumptions towards audiences and usage scenarios, and considering the politics of cultural collections and their interfaces more broadly.
\section{Background}
If Nowviskie explicitly identifies the problem with current digital libraries as a design problem~\cite{nowviskie2016}, it is because the
design of interfaces for cultural collections frames cultural materials typically as passive representatives of the past and orients audiences toward specific modes of receptivity and praxis. The design challenge for digitized collections then is at least two-fold: (1)~to find ways to better represent the materials themselves, and (2)~to re-orient audiences toward new modes of active embodied engagement. 

With Digital Bricolage we work toward embodied approaches to cultural collections, because we recognize (as others have) that any material artifact has multiple dimensions—--including a performative dimension--—and that human beings make meaning in part by actively interacting with the material world from situated perspectives. Researchers from the humanities (e.g., Drucker~\cite{drucker2002}) as well as from HCI (e.g., Giacardi \& Karana~\cite{Giaccardi_2015}; Dourish~\cite{dourish2004}) have drawn on phenomenological theories to emphasize both the performative dimension of materiality and to highlight how human beings make meaning or interpret the world by interacting with it in embodied ways (rather than encountering it passively or abstractly). 

While dominant modes of digital display tend to emphasize the semantic content of digitized print artifacts while obscuring their material (and performative) dimensions, we see opportunities to recreate this dimension in digital spaces with the goals of enhancing processes of meaning-making that depend on material engagement and supporting a more robust appreciation of the complexity of print records. Below we outline how Nowviskie's identified challenges were addressed by previous work, before introducing the canonical interfaces we asked participants to consider in comparison to Digital Bricolage.

\paragraph{Alternative Modes of (Visual) Engagement.}
A number of approaches have sought to re-orient audiences toward alternative, more exploratory modes of engagement with digital collections. Dörk et al.’s concept of the \textit{``information flâneur''} offers a counterpoint to goal-driven search, framing information seeking as a pleasurable, wandering experience through digital space~\cite{dork2011}. One of the most influential contributions in this space is Whitelaw’s concept of \textit{``Generous Interfaces,''} which encourages open-ended browsing by displaying the richness of a collection---often through arrays of thumbnails, while still supporting targeted search~\cite{Whitelaw_2015}. This ethos is echoed in various visualization techniques that integrate thumbnail images as glyphs within charts, clusters, or timelines~\cite{bowkerstar1999,emdialog,manovich2012}, allowing visual aesthetics and metadata to co-exist. Thudt et al. further extend this approach by proposing visualizations that combine content-related metadata with the visual and material characteristics of cultural artifacts (e.g., color, size), enabling serendipitous discoveries and more open-ended exploration~\cite{Thudt_2012}.

These approaches illustrate how visual aspects of a collection offer a multiplicity of entry points and showcase a kind of visual generosity that can highlight the breadth and uniqueness of a collection. However, alternative modes of visual engagement, let alone other material features of collected artifacts, are rarely explored in canonical interfaces to digital collections. 

\paragraph{Canonical Interfaces to Digital Collections.}
Digital (online) platforms of libraries, archives or museums have started to move beyond textual search interfaces toward more sophisticated visual
interfaces that allow for both targeted and exploratory search. However, approaches to interface design typically do not move beyond showcasing digitized artifacts in the form of thumbnail images. We illustrate this based on three examples that, we argue, dominate the field in shaping and controlling access to a number of major print-based collections and that form the backdrop to exploring more speculative approaches as illustrated by Digital Bricolage. 

\begin{figure}[b!]
\centering
  \includegraphics[width=4.5cm]{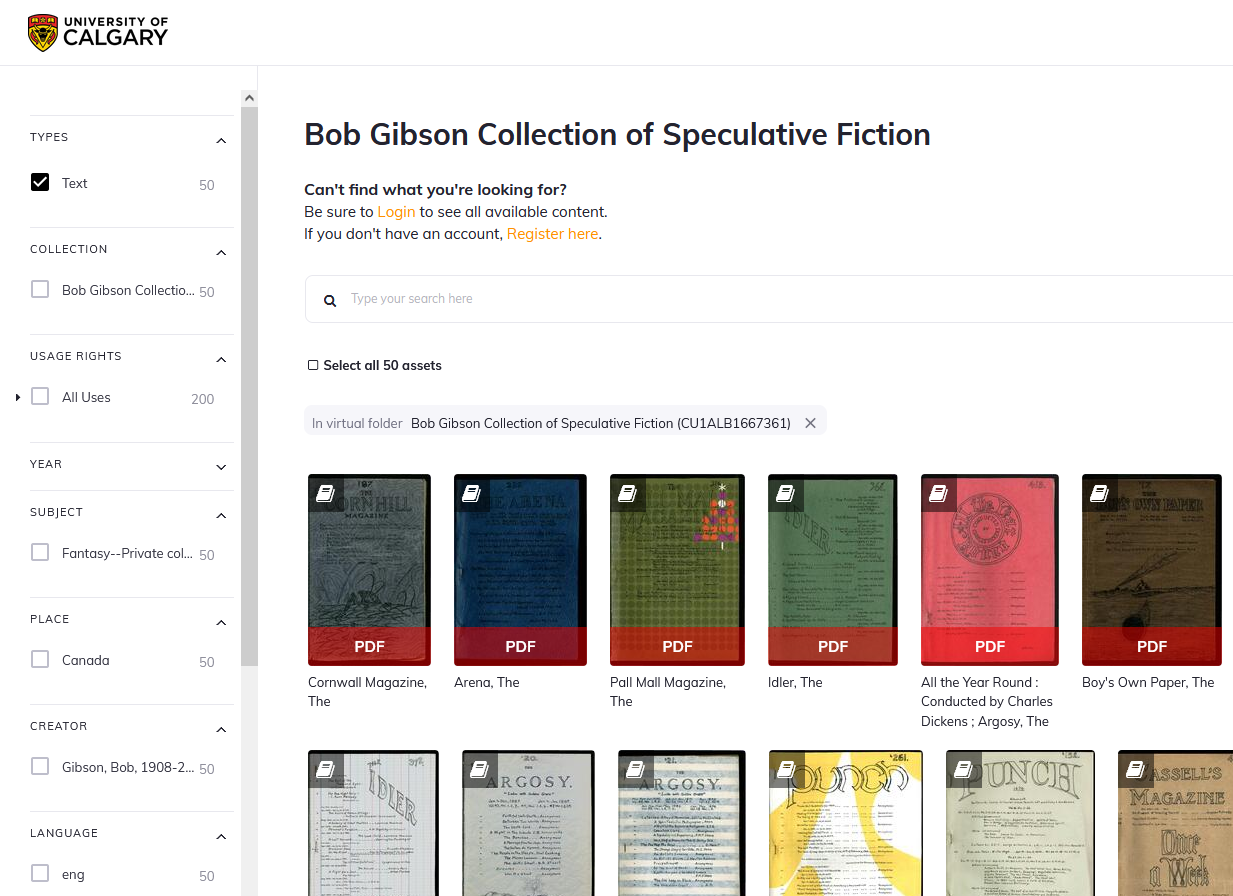}
  \includegraphics[width=3.5cm]{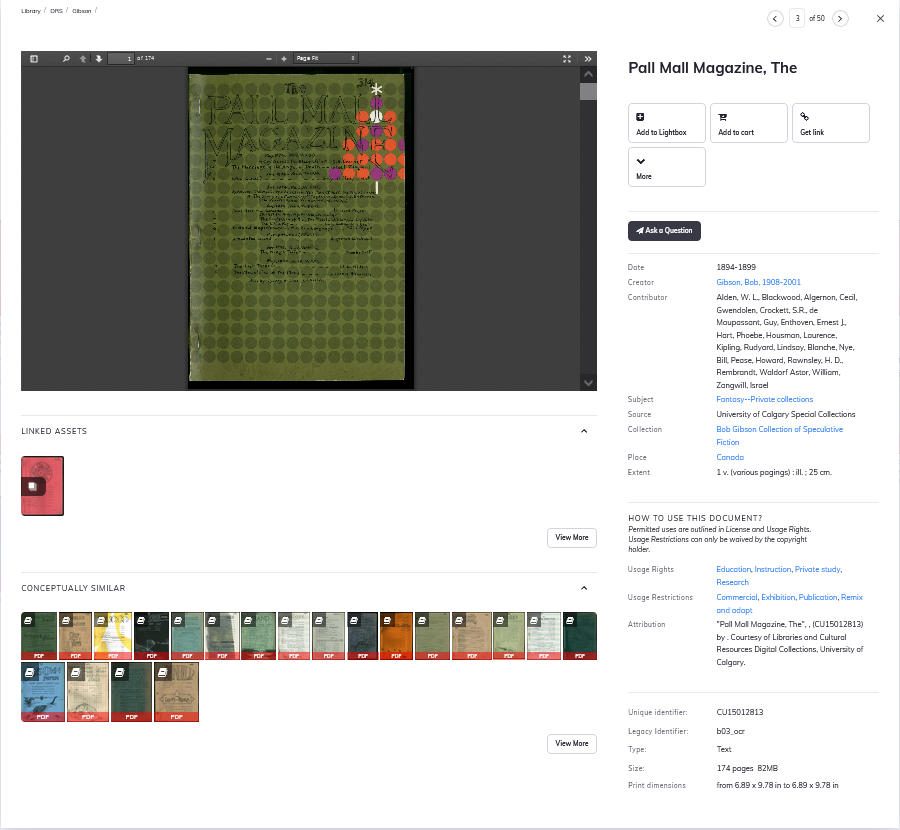}
  \caption{Typical thumbnail view of the Bob Gibson Anthologies of Speculative Fiction in ContentDM.}
  \label{fig:gibson}
\end{figure}

\begin{figure}[htbp]
\centering
  \includegraphics[width=8cm]{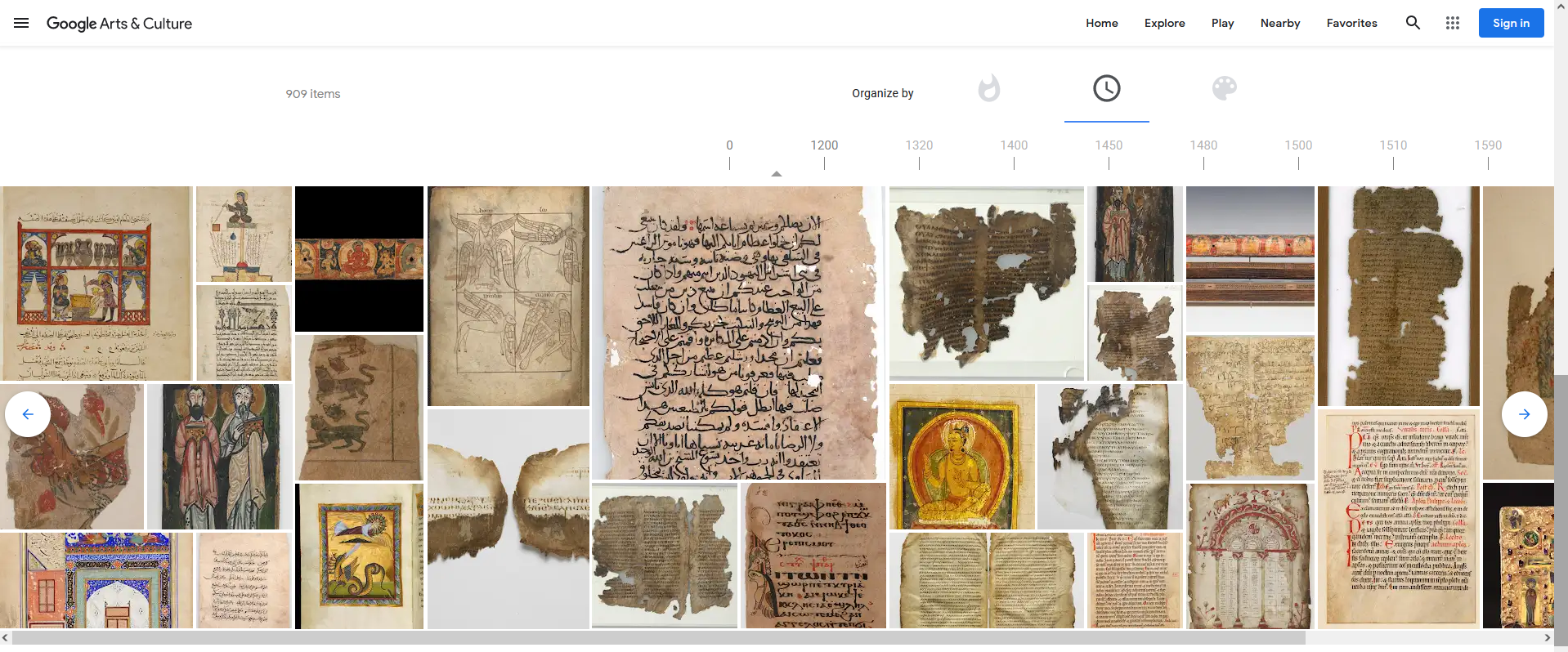}
  \caption{Overview screen in Google Arts and Culture of print manuscripts organised on a timeline.}
  \label{fig:google}
\end{figure}

Interfaces for digitized print-based collections commonly feature landing pages with grid-based thumbnail views (see Fig.~\ref{fig:gibson} \&~\ref{fig:google}). Provided filters are typically textual, with occasional inclusion of visual metadata such as colour. Clicking a thumbnail opens a detail page for deeper exploration. Platforms such as CONTENTdm\footnote{Viewable at: \url{https://www.oclc.org/en/contentdm.html}}, a standard system for managing and making available digital collections of cultural heritage materials, but also the Digital Bodleian\footnote{Viewable at: \url{https://digital.bodleian.ox.ac.uk/}}, a digital platform to make treasures the Bodleian Libraries and Oxford college libraries explorable, exemplify this structured approach to digital cultural heritage (see Fig.~\ref{fig:gibson}). In contrast, the interfaces of the Google Arts \& Culture initiative\footnote{Viewable at: \url{https://artsandculture.google.com/}} employs a more fluid layout. While still dominated by thumbnails, its `ribbon' interface allows endless, less structured browsing (see Fig.~\ref{fig:google}), combining content-based filters with social ones, such as popularity.

Recent works of our own~\cite{mining} and others~\cite{glinka2017,gortana2018} that have started to question and explore alternative modes of engagement with digitized collections. Driven by an urge to break away from canonical interfaces, we were curious to experiment with juxtaposing material and content-based qualities of print-based cultural collections, and to engage others in re-thinking interfaces in this context.
\section{Digital Bricolage}
What would it mean to imagine digital collections as performance spaces? What opportunities for embodied experience might we discover if we embrace the material qualities of print-based collections when designing interfaces for digital access? With these questions in mind, we designed Digital Bricolage for the Gibson Anthologies of Speculative Fiction---a unique collection of hundreds of handcrafted ``fanzines'' put together by the science fiction collector and fan artist Bob Gibson. The collection contains thousands of science fiction stories from the 1840s to the 1990s~\cite{forlini2016,specpractices}. The Gibson Anthologies are an exemplary collection where content, material and visual qualities are uniquely interwoven, and, as such, they beg for a more tailored approach to interface design that honours not only content, material and visual qualities of the collection but is also mindful of the process in which items were created (manually, using scraps carefully curated by Gibson as a kind of citizen scholar). 

Digital Bricolage (developed in Axure RP 10) is the result of an iterative and interdisciplinary design process that included sketching to explore and iteratively refine initial design concepts. From there, low fidelity prototypes were developed and discussed within our research team which included interface design experts as well as a researcher in literary studies who works closely with the Gibson Anthologies. Digital Bricolage can be considered an experimental and speculative prototype that illustrates central ideas, in order to inform a dialogue around these. Our design process was driven by the urge to juxtapose material qualities and content of digitized print-based collections, moving away from grid-based interfaces, and toward alternatives that might align more closely with our in-person experiences of and engagement with physical collections. Our goal was to create a visual and visualization-driven interface that would build upon and leverage real-world interactions when appropriate, and that surpassed them when possible, combining imagined, desired/speculative, and remembered experience.
It was important to us to balance a sense of familiarity with visual, interactive novelty by using visual analogies from the physical world in combination with abstract forms drawing from data visualization. 

Digital Bricolage juxtaposes material qualities and content of digitized print-based collections by providing unique spatial layout and filter options driven by material aesthetics, alongside elements indicating a sense of physical scale. Emphasizing the unique physical, material, and visual qualities of the Gibson Anthologies as a print-based, handcrafted collection, we attempted to amalgamate
analogue and digital aesthetics, acting as a subtle nod to the possibilities opened up by re-presenting print-based collections in a digital environment. We present individual features of Digital Bricolage below; a video is provided as supplementary material.

\paragraph{Spatial Layout.}
All interactions in Digital Bricolage are centred around an interactive canvas---the collection canvas---where the anthologies can be arranged in a pile- or shelf-based layout (see Fig.~\ref{fig:teaser}.1~\&~1.2); each emphasizing different aspects and allowing for different modes of exploration. 

The \textit{\textbf{shelf-based layout}} arranges anthologies side-by-side, akin to a physical bookshelf. Anthologies are represented through extracts of their covers, highlighting the anthology's title, the typography used, cover illustrations, and cover paper type. The anthology's size on the shelf represents the physical height and page count of their print-based counterpart. The shelf-based layout facilitates familiar search and explorations such as going through titles one-by-one on a physical bookshelf. Tooltips augment the spines on the shelf by providing a preview of anthology covers' illustrations.

In contrast, the \textit{\textbf{pile-based layout}} invites visitors to freely group and (re-)arrange anthologies. It, thus, departs from the canonical, static grid-based view typical in digital interfaces to cultural collections and deliberately supports non-linear explorations. The pile-based layout intentionally embraces messiness and disorganization, evocative of the bricolage of the collection itself and the possibilities it presents for audiences. Relative size of items is preserved in the thumbnail images. The lack of content-based ordering indirectly emphasizes physical and visual features of individual items, such as size, cover colour, and cover illustrations. As shown in Figure~\ref{fig:teaser}.5, a tooltip provides information about an anthology's material (cover and page material, binding) and visual features (i.e., the most dominant colours). 

Anthologies can be automatically arranged into smaller piles (e.g., based on their colour; Fig.~\ref{fig:teaser}.3) to facilitate a more structured exploration while still allowing for custom arrangements.

\paragraph{Representing Physical Scale.}
It was important to us to provide a relative sense of physical scale at all times in Digital Bricolage, which is often lost in canonical interfaces where representations of digitized items are homogenized. Anthologies in the collection canvas are therefore always shown to relative scale. In addition, a scale measure to the right of the canvas (see Fig.~\ref{fig:teaser}.1~\&~1.2) hints at the actual size of anthologies. It provides an important orientation when zooming into the collection canvas and can also be used directly for zooming.

\paragraph{Visual \& Material Filters.}
Digital Bricolage provides several ways of searching and filtering a collection, bridging its content and its visual and material qualities.

\textit{\textbf{Cover Colour \& Publication Year.}} A timeline shows the publication years of science fiction stories (squares) within their individual anthology (square colour; see Fig.~\ref{fig:teaser}.8). It acts as a filter by publication year---selecting individual years or time ranges provides the corresponding stories and their anthologies. However, the timeline is also linked to anthology covers and their colours. Colour patches emit a semi-transparent triangle to highlight the publication years an anthology spans based on the stories included, thus, juxtaposing publication year ranges with associated cover colours.

\textit{\textbf{Colour \& Texture.}} A colour filter allows searching for anthologies by cover colour (see Fig.~\ref{fig:teaser}.7); a colour wheel presents filter choices in a fluid way. `Sample squares' can be created and adjusted within the wheel to select the desired colour. The sample squares present the selected colour directly through the cover material of the corresponding anthologies, also providing a glimpse of surface texture. Multiple colour selections are possible, resulting in a logical `OR' combination of colour filter criteria. For example, creating a red and blue sample square results in anthologies with dominant colours red or blue in the collection canvas (see Fig.~\ref{fig:teaser}.6).

\textit{\textbf{Physical \& Content Size.}} Anthologies can be filtered based on their physical size and the number of stories included (Fig.~\ref{fig:teaser}.9). Anthology sizes are clustered into four distinct categories represented by four rectangles of corresponding width and height dimensions. The darker the rectangle the more anthologies are close to these dimensions. The bars below the cover dimensions represent the average number of stories included in the selected anthologies.

All filtering options above can be combined and filter results are always visible for further exploration in the collection canvas.

\section{Expert Reflections on Digital Bricolage}
We created Digital Bricolage, in part, to explore our own ideas of what it could mean to embrace the material qualities of print-based collections when designing interfaces for digital access. However, Digital Bricolage also created an opportunity to start a dialogue with others on (re-)imagining digital collections as performance spaces that can challenge the experiences that dominant platforms currently invite and/or discourage. We used Digital Bricolage as a visualization design probe~\cite{wallace2013} to explore perspectives on digital cultural collections with 8 creative professionals (graphic/product designers, writers) and 15 collection experts (librarians, archivists, curators). Each of the 23 individual 1.5-hour interviews began with an unaided, unguided, think-aloud exploration of three existing interfaces (a ContentDM-based digital library, the Digital Bodleian, and Google Arts \& Culture) focusing on aspects that sparked curiosity, worked well, or posed pain points or limitations. Participants then watched a video walkthrough of Digital Bricolage and were encouraged to pause the video and offer immediate thoughts and feedback when relevant. We followed with questions to elicit participants' perspectives on the observed features and their potential applicability to other cultural collections. The video and interview guide are available as supplementary materials. Below, we summarize responses to Digital Bricolage in relation to comments on the initial three interfaces.

\subsection{Reactions to Canonical Interfaces}
The image-heavy visual experience of the \textbf{\textit{Google's Arts \& Culture}} platform combined with the breadth of information available, provided some participants with an immersive and generous experience but overwhelmed others. Participants found the platform to be most suitable for curiosity-driven browsing without deep-dives. Overall, the interface reminded participants of a museum and was able to trigger sensory and embodied experiences in some cases, but this was heavily dependent on the item being viewed and was undermined by the disturbing lack of transparency with regards to curatorial choices and the sources of items displayed.

Participants perceived \textbf{\textit{CONTENTdm}} as a more serious, content-oriented approach to cultural collections. While some found it dry and outdated, others experienced it more accessible and structured. Most agreed that the platform felt like it belonged to a university or library, intended to be used by researchers and experts due to its exhaustive and systematic nature. However, participants also felt that the interface lacked context and discoverability, relying on a certain level of user expertise on the presented items. Overall, the panning and zooming features paired with the physical features of items visible, such as the bindings, allowed users to get a sense of items' material ``feel''. That being said, for some participants even small mismatches between the forms of interactions on the platform and real-world interactions (e.g., scrolling down instead of turning pages left and right) prevented a more immersive experience.

Finally, the \textbf{\textit{Digital Bodleian}} was positively received as being well-balanced in terms of content and visual information. The high-resolution images and zooming capabilities led to an immersive experience. Much like Google Art and Culture, some participants found the interface to be overwhelming or distracting due to its image-heavy nature, but also because of its high customizability, breadth of content, and the rigorous scholarly information provided. Nevertheless, participants felt that the interface was suitable for both targeted exploration and for casual browsing or exploring a ``big universe" of items. Much like CONTENTdm's Gibson Anthologies, the ability to zoom into the high-quality scans of items provided a tactile sense of the items and allowed a focus on material aspects that felt like ``visually [touching]" the items. 

These findings show that in all three platforms, links to physical experiences are triggered by the presence of high-quality images that can account for traces of production and handling, blemishes, and unique material qualities of the item, paired with pan \& zoom features to explore these images in detail. This means that participants' focus on and exploration of materiality was heavily dependent on (1) the item itself (i.e., presence of interesting visual and material features), (2) how it was digitized (i.e., definition of digital images), and (3) the platform's interactive features supporting exploration of the item. 

These findings are interesting to consider in the context of how the same participants responded to the idea of Digital Bricolage which intentionally presents visual and material features of print-based cultural collections in different ways. 

\subsection{Reactions to Digital Bricolage}
Participants' initial reactions to Digital Bricolage's information and interface design were mostly positive: they described it as playful, fun, intuitive, inviting and friendly. Below we summarize some of their more salient responses, using ``[C]'' for statements from creative professionals, and ``[E]'' for collection experts. 

\paragraph{Modes of Exploration \& Spatial Arrangement.} Participants appreciated how the multiplicity of exploration options and flexibility in arrangements can allow for self-curation, providing a sense of agency and control over the experience: \textit{``I have an investment in that, it becomes mine, [... even though] you've given me the content, [...] I'm more engaged with it... There's an investment in that which there isn't in the other ones [digital interfaces]"}~[C9].

The ability to spatially re-organize items appeared to offer a way of engaging with items which resonated strongly with participants: \textit{``I really like the piling up, and then you can kind of move around. I would even like to label them, you know, so you kind of put the pile there and you can kind of write something underneath, and then it works like a working archival [...] table"}~[E8];  \textit{``Moving them around in piles and the manipulation... I really like that because, that's what I wish I could do in special collections”}~[E3]. 

The move away from the traditional grid layout to present collections stimulated a desire to learn more about the collection: \textit{``There are certain aspects of viewing the collection or groups of items in that way that sort of stimulates your mind and your experience in a different way than seeing it in a grid. It kind of gives you a little bit of that spark where you think I wanna know more”}~[E12]. 

\paragraph{Balancing Physical and Digital Affordances.} Participants felt that Digital Bricolage can invoke explorations familiar from the physical world, enabling sensemaking of \textit{``the ways that these materials have been presented”}~[C5]. Another participant speculated that it was this familiarity paired with the digital augmentation (e.g., having tooltips showing details of anthologies in the shelf layout) that can work well: \textit{``a bit of a surprise element, but it's still a familiar setting. I really like that”} [C9]. 

\paragraph{Visual \& Material Understanding.} When participants spoke about the collection presented in Digital Bricolage, it was the aesthetic and material aspects they focused on, rather than textual content within the individual artifacts. This suggests that Digital Bricolage encourages a shift from the content-oriented mindset and toward an appreciation of the physicality and aesthetic aspects of a collection and its items: \textit{``This strikes me as a very book history approach to digitizing materials. [...] you can start to see that this is the sort of physical object, actual books, not just holders of text. I think that's really exciting. I think that offers a lot of opportunities for people to approach material in different ways”}~[E10].

\paragraph{Links to Embodied Experience.}
Some participants' responses and reactions indicate that Digital Bricolage allows links to tactile and physical experiences. For example: \textit{``I thought it [Digital Bricolage] gives me a similar experience as if I'm [in a] library''}~[C1]. Bringing out the material aspects seems to encourage the imagined tactile experience: \textit{``I like this idea of being able to explore by texture, that's not necessarily something I think you would typically use as a way to get into a particular collection on a digital interface now, but it might be something that I would do in real life whatever the context”}~[C5]; \textit{``Having that touch information, that tactile information, helps people realize that this exists in real life, like its not just a scan on a screen, it's a real thing that you could go pick up. And that's information that's not really in catalog records, that's not in most digital interfaces"}~[E6].

Notably participants felt Digital Bricolage highlighted the three-dimensionality of items: \textit{``You do get an idea [...] much better about the three dimensionality, besides the comparison to another, I mean, if you imagine you had some books and pamphlets [...] and you can straight away see their volume in relation to each other”}~[E8].

Finally, Digital Bricolage led some participants to challenge their ways of working with current resources: \textit{``it makes you re-evaluate how you currently use resources [...] just because it's the way that we did it before, doesn't mean it's the way we should continue to do it... this gives you that opportunity to think outside `what is'”}~[C7].

\section{Discussion}
\label{sec:figure_credits_inst}
Our work on Digital Bricolage is speculative and exploratory in nature, resulting in a number of open questions which we raise and contribute towards, but which we also invite future research to explore and carry forward. 

\textbf{\textit{How can we better represent the materiality of digitized print collections?}}
Historically, the physical aesthetics of print collections have been largely treated as \textit{``window-dressing''} [p.14]~\cite{mining} or \textit{``an after-thought''} [p.5]~\cite{sandcastles}. Digital Bricolage was inspired by previous work exploring (i) non-linear, dynamic visual metaphors to support serendipitous discovery, systematic search, and interpretive flexibility \cite{gortana2018} and (ii) the use of alternative filtering mechanisms, such as temporal visualisations, textural clustering, and theme-based filtering to encourage slow exploration and reflective engagement \cite{glinka2017}. We aim to showcase and celebrate collection items’ materiality by, for example, enabling filtering and the exploration of collection items based on their visual and material qualities, and providing a sense of relative and absolute scale. Participants appreciated these features and commented on the improved sense of tactile materiality in Digital Bricolage as compared to the other canonical interfaces. However, our work thus far just scratches the surface when it comes to opportunities for representing materiality of digitized print collections. Marrying advances in data visualization with tangible interaction or augmented and virtual reality are just a few avenues to explore in the future.

\textit{\textbf{How can we move towards contextual and performative generosity?}}
Generous interfaces, as described by Whitelaw~\cite{Whitelaw_2015}, tend to focus on visual generosity. Our participants associated generous interfaces with curiosity-driven open-ended browsing (see Google Arts \& Culture). By contrast, they compared interfaces allowing for research and more in-depth exploration and information seeking to a \textit{``closed window''} or to being \textit{``pigeon-holed''}. 

With Digital Bricolage, we sought to move beyond visual presentation toward contextual and performative generosity, emphasizing not only the relationships between items but also users’ potential to engage with collections as performance spaces. This led participants to consider diverse exploration modes: the bookshelf view was seen as conducive to focused searching and conveyed a sense of physical scale, while the book pile view suggested open-ended browsing and serendipitous discovery. In both layouts, items were presented collectively rather than in isolation, encouraging participants to muse connections and appreciate the collection as a meaningful whole. Participants noted that Digital Bricolage evoked the sense of items occupying and belonging to a physical space. This contrasted with Google Arts \& Culture, where items felt detached from their collections, and with ContentDM, where grid layouts were seen as flattening physical and contextual uniqueness.

Some participants expressed a desire for an even stronger connection to real items being stored in a real physical location: \textit{``it would be nice to see images of these things sort of as they are stored in their place''}~[E8], something that can guide future explorations of effective interfaces for print-based digital collections.

\textbf{\textit{How can we elicit active embodied engagement, as opposed to passive consumption?}}
Participants described feeling like passive observers in existing interfaces—where interactions often clashed with their embodied experiences of print collections (e.g., vertical scrolling in the Digital Bodleian versus turning pages in a book). In response to Nowviskie’s calls, Digital Bricolage introduces flexible layouts that support playful, interactive engagements grounded in familiar physical metaphors, such as rummaging through books or browsing books on shelves. As a result, participants speculated that Digital Bricolage could be not merely a `resource' but a research `tool,' suggesting a shift toward active, exploratory use.

This sense of engagement seems to be tied to the emphasis of tactile, performative interaction—particularly through self-curation, which invited participants to muse about agency and ownership. While self-curation is not new to digital collections, participants valued being able to arrange materials themselves in meaningful ways. By contrast, participants viewed existing interfaces as exerting control. The polished curation of Google Arts \& Culture, for instance, limited their ability to shape their browsing experience. ContentDM's institutional framing (as a university or library interface) also reinforced a sense of top-down authority, amplified by a lack of transparency about what is included or omitted, making visible the power embedded in its structure.

Although Digital Bricolage may not fully achieve Nowviskie’s vision of speculative collections in terms of inviting play to imagine new narratives that confront past and present injustices, it marks a step toward that goal. By adopting the aesthetic identity of the Gibson collection, Digital Bricolage highlights the idiosyncrasies of a `bricolage' archive assembled by an `outsider' to institutional power, a citizen scholar and fan artist whose lifelong engagement with science fiction challenges dominant narratives of the genre~\cite{forlini2022}. Digital Bricolage draws on this to prompt participants to imagine their own speculative curations and contribute to ongoing conversations about interface design for print-based cultural heritage. Future work should continue exploring embodied engagement not as an end in itself, but as a means to surface alternative narratives, counter power imbalances, and foster active, collaborative, and creative participation over passive consumption~\cite{lee2020}.

\acknowledgments{
The authors wish to thank all the interview participants and the University of Calgary's Libraries and Cultural Resources who provided the images from the Gibson Anthologies collection.}



\bibliographystyle{abbrv-doi}

\bibliography{template}

@misc{nowviskie2016,
  author = {Nowviskie, B.},
  title = {Speculative Collections [Blog Post]},
  year = {2016},
  url = {https://nowviskie.org/2016/speculative-collections/},
  lastaccessed = {2024},
}

@article{digitalart,
author = {Marcos, A. F. and Branco, P. and Carvalho, J. A.},
doi = {10.4018/9781605663524.ch001},
journal = {Handbook of Research on Computational Arts and Creative Informatics},
number = {April 2017},
title = {The computer medium in digital art's creative process},
year = {2011}
}

@INBOOK{drucker2002,
 author = {Drucker, J.},
 title = {Intimations of Immateriality: Graphical Form, Textual Sense, and the Electronic Environment},
 booktitle = {Reimagining Textuality: Textual Studies in the Late Age of Print},
editor = {Loizeaux, E.B and Fraistat, N.},
 pages = {152--177},
 publisher = {Madison: University of Wisconsin Press}
}

@article{facet,
author = {Koren, J. and Zhang, Y. and Lue, X.},
journal = {Proceedings of the 17th International Conference on World Wide Web, WWW},
title = {Personalized Interactive Faceted Search},
year = {2008},
doi = {10.1145/1367497.1367562 }
}

@article{visualizing,
author = {Mer\v{c}un, T. and \v{Z}umer, M.},
journal = {Proceedings of the Library In the Digital Age},
pages = {104--115},
title = {Visualizing for explorations and discovery},
year = {2010}
}

@article{mandala,
author = {Brown, S. and Ruecker, S. and Antoniuk, J. and Farnel, S. and Gooding, M. and Sinclair, S. and Gabriele, S.},
number = {1},
volume = {2},
journal = {Digital Studies/le Champ Numérique},
title = {Reading {Orlando} with the {Mandala Browser}: A Case Study in Algorithmic Criticism via Experimental Visualization},
year = {2011}
}

@article{emdialog,
author = {Hinrichs, U. and Schmidt, H. and Carpendale, S.},
journal={IEEE Transactions on Visualization and Computer Graphics}, 
title={{EMD}ialog: Bringing Information Visualization into the Museum}, 
year={2008},
volume={14},
number={6},
pages={1181-1188}
}

@book{bowkerstar1999,
author = {Bowker, G. C. and Star, S. L.},
year = {1999},
title = {Sorting Things Out: Classification and Its Consequences},
publisher = {Cambridge, Massachusetts: MIT Press}
}

@article{specpractices,
author = {Hinrichs, U. and Forlini, S. and Moynihan, B.},
doi = {10.1109/TVCG.2015.2467452},
issn = {10772626},
journal = {IEEE Transactions on Visualization and Computer Graphics},
number = {1},
pages = {429--438},
title = {Speculative practices: Utilizing {InfoVis} to explore untapped literary collections},
volume = {22},
year = {2016}
}

@article{forlini2016,
author = {Forlini, S. and Hinrichs, U. and Moynihan, B.},
year = {2016}, 
title = {The Stuff of Science Fiction: An Experiment in Literary History},
journal = {Digital Humanities Quarterly (DHQ)},
volume = {10},
number = {1},
}

@article{lee2020,
author = {Li, J. and Wu, L. and Hong, R. and Zhang, K. and Ge, Y. and Li, Y.},
title = {A Joint Neural Model for User Behavior Prediction on Social Networking Platforms},
year = {2020},
issue_date = {December 2020},
publisher = {Association for Computing Machinery},
address = {New York, NY, USA},
volume = {11},
number = {6},
issn = {2157-6904},
url = {https://doi.org/10.1145/3406540},
doi = {10.1145/3406540},
journal = {ACM Trans. Intell. Syst. Technol.},
month = {sep},
articleno = {72},
numpages = {25}
}

@article{forlini2022,
  author = {Forlini, S.},
  title = {Periodical Speculations: Early "Science Fiction" and Popular Victorian Weeklies},
  journal = {Science Fiction Studies},
  volume = {49},
  number = {1},
  pages = {1--31},
  year = {2022},
  doi = {10.1353/sfs.2022.0003},
  publisher = {SF-TH Inc.},
  url = {https://muse.jhu.edu/article/853494}
}

@article{sandcastles,
author = {Hinrichs, U. and Forlini, S. and Moynihan, B.},
doi = {10.1093/llc/fqy051},
issn = {2055768X},
journal = {Digital Scholarship in the Humanities},
pages = {I80--I99},
title = {In defense of sandcastles: Research thinking through visualization in digital humanities},
volume = {34},
year = {2019}
}

@inproceedings{wallace2013,
  title={Making design probes work},
  author={Wallace, J. and McCarthy, J. and Wright, P. C. and Olivier, P.},
  booktitle={Proceedings of the SIGCHI Conference on Human Factors in Computing Systems},
  series={CHI '13},
  pages={3441--3450},
  year={2013},
  publisher={ACM},
  doi={10.1145/2470654.2466473}
}

@Article{manoff2006,
    author="Manoff, M.",
    journal="Libraries and the Academy", 
    title="The Materiality of Digital Collections: Theoretical and Historical Perspectives", 
    year="2006",
    pages="311--325",
    volume="6",
    number="3",
    doi="https://doi.org/10.1353/pla.2006.0042"
}

@article{drucker2013,
author = {Drucker, J.},
year = {2013},
title = {Is There a ``Digital'' Art History?},
journal = {Visual Resources: An International Journal of Documentation},
volume = {29},
number = {1-2},
pages = {5--13}
}

@article{mining,
author = {Forlini, S. and Hinrichs, U. and Brosz, J.},
doi = {10.16995/olh.282},
issn = {2056-6700},
journal = {Open Library of Humanities},
number = {2},
pages = {1--36},
title = {Mining the material archive: Balancing sensate experience and sense-making in digitized print collections},
volume = {4},
year = {2018}
}

@inproceedings{Thudt_2012,
  author = {Thudt, A. and Hinrichs, U. and Carpendale, S.},
  title = {The Bohemian Bookshelf: Supporting Serendipitous Book Discoveries through Information Visualization},
  booktitle = {Proceedings of the 2012 SIGCHI Conference on Human Factors in Computing Systems (CHI ’12)},
  pages = {1461--1470},
  year = {2012},
  doi = {10.1145/2207676.2208607},
}

@Article{Whitelaw_2015,
author = "Whitelaw, M.",
number = "1",
pages = "1--16",
title = "Generous Interfaces for Digital Cultural Collections",
journal = "DHQ : Digital Humanities Quarterly",
volume = "9",
year = "2015",
}

@article{stauffer2012,
author = {Stauffer, A.},
year = {2012}, 
title = {The Nineteenth-Century Archive in the Digital Age},
journal = {European Romantic Review},
volume = {23},
number = {3},
pages = {335--341},
doi = {10.1080/10509585.2012.674264}
}

@book{dourish2004,
author = {Dourish, P.},
year = {2004},
title = {Where the Action Is: The Foundations of Embodied Interaction}, 
publisher = {MIT Press}
}

@inproceedings{Giaccardi_2015,
    author = {Giaccardi, Elisa and Karana, Elvin},
    title = {Foundations of Materials Experience: An Approach for HCI},
    year = {2015},
    isbn = {9781450331456},
    publisher = {Association for Computing Machinery},
    address = {New York, NY, USA},
    url = {https://doi.org/10.1145/2702123.2702337},
    doi = {10.1145/2702123.2702337},
    booktitle = {Proceedings of the 33rd Annual ACM Conference on Human Factors in Computing Systems},
    pages = {2447–2456},
    numpages = {10},
    location = {Seoul, Republic of Korea},
    series = {CHI '15}
    }

@inproceedings{dork2011,
  author = {Dörk, M. and Carpendale, S. and Williamson, C.},
  title = {The Information Flaneur: A Fresh Look at Information Seeking},
  booktitle = {Proceedings of the 2011 Annual Conference on Human Factors in Computing Systems (CHI '11)},
  pages = {1215--1224},
  year = {2011},
  doi = {10.1145/1978942.1979124},
}

@article{glinka2017,
  author      = {Glinka, K. and Pietsch, C. and Dörk, M.},
  title       = {Past Visions and Reconciling Views: Visualizing Time, Texture and Themes in Cultural Collections},
  journal     = {Digital Humanities Quarterly (DHQ)},
  volume      = {11},
  number      = {2},
  year        = {2017},
}

@article{gortana2018,
  author      = {Gortana, F. and von Tenspolde, F. and Guhlmann, D. and Dörk, M.},
  title       = {Off the Grid: Visualizing a Numismatic Collection as Dynamic Piles and Streams},
  journal     = {Open Library of Humanities},
  volume      = {4},
  number      = {2},
  pages       = {1--25},
  year        = {2018},
  doi         = {10.16995/olh.280},
}

@article{manovich2012,
author = {Manovich, L.},
journal = {Abstartion Now},
title = {Media Visualization: Visual Techniques for Exploring Large Media Collections},
year = {2012},
}
\end{document}